\documentclass[12pt,a4paper]{article}

\usepackage{amsmath, amssymb, amsthm}
\usepackage{mathtools}
\usepackage{geometry}
\usepackage{natbib}
\usepackage{hyperref}
\usepackage{booktabs}
\usepackage{bm}
\usepackage{graphicx}
\usepackage{pgf}
\usepackage{enumitem}
\usepackage{placeins}
\usepackage{xcolor}
\usepackage{soul}
\usepackage[framemethod=TikZ]{mdframed}

\theoremstyle{definition}
\newtheorem{remark}{Remark}

\DeclareMathOperator{\E}{\mathbb{E}}

\title{Bayesian Seasonal Adjustment for Survey Time Series}

\author{Siu-Ming Tam\\
Tam Data Advisory, Australia}

\date{\today}

\begin{document}
\maketitle

\begin{abstract}
Seasonal adjustment procedures used by national statistical offices---X-11
and X-12-ARIMA---treat each survey estimate as an exact observation,
discarding the accompanying standard errors that survey methodologists
routinely compute.
This paper closes that gap by embedding time-varying sampling error variances
into a Basic Structural Model (BSM), extending a recently proposed
Dynamic Mini-Max (DMM) Bayesian framework for survey estimation.
Via the Harvey--Todd equivalence, BSM with zero measurement error variance
reduces to X-11-style seasonal adjustment, so DMM-BSM is a principled
Bayesian generalisation of existing practice rather than a departure from it.

A two-block Gibbs sampler delivers the full joint smoothing posterior of
the latent state trajectory.
Exact credible intervals for the trend level, $k$-step trend movements
($k=1,2,\ldots$), and seasonally adjusted estimates follow directly,
together with posterior probabilities of directional change---outputs
that X-11 cannot provide.

Simulation studies with within-replication parameter estimation confirm
substantially higher credible interval coverage than the X-11-equivalent
model across both large and small survey domains.

Applied to 120 months of Australian Bureau of Statistics Labour Force
Survey data, once sampling variance is modelled the maximum likelihood
estimate of stochastic seasonal variance collapses to zero---evidence
that apparent seasonal fluctuations in the published series are largely
attributable to measurement noise rather than genuine seasonal drift,
a distinction X-11 cannot make.
\end{abstract}

\section{Introduction}
\label{sec:intro}

National statistical offices (NSOs) routinely publish two types of labour
force estimates: levels at each wave and movements between waves. The X-11
seasonal adjustment procedure \citep{shiskin1967} and its successor
X-12-ARIMA \citep{findley1998} are the standard tools used worldwide to
decompose survey time series into trend, seasonal, and irregular components
before publishing these estimates. Despite their widespread adoption, these
procedures share a structural limitation: they treat each survey estimate
$Y_t^d$ as an exact observation, making no use of the accompanying standard
error $\mathrm{SE}_t^d$ that survey methodologists routinely compute alongside
every published figure. As a consequence, a month with a coefficient of
variation (CV) of 3\% and a month with a CV of 0.5\% receive the same weight
in the seasonal filter, even though the former has a standard error six times larger (and a sampling
variance thirty-six times larger) than the latter.

The theoretical connection between X-11 and model-based seasonal adjustment is
well established. \citet{harveytod1983} showed that the reduced form of a
Basic Structural Model (BSM) with a local level trend and a stochastic seasonal
component is the airline ARIMA$(0,1,1)(0,1,1)_{12}$ model. \citet{clevelandtiao1976}
showed that X-11 seasonal adjustment is asymptotically equivalent to signal
extraction from the airline model; issues in model identification for seasonal adjustment filters are analyzed by \citet{bell1984}. Together, these results form an equivalence
chain:
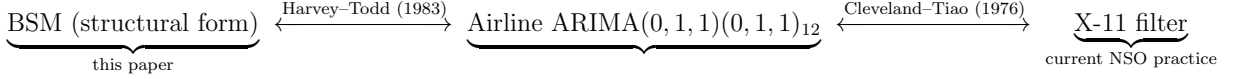
\begin{equation*}
\resizebox{\textwidth}{!}{$
    \underbrace{\text{BSM (structural form)}}_{\text{this paper}}
    \;\xleftrightarrow{\text{Harvey--Todd (1983)}}\;
    \underbrace{\text{Airline ARIMA}(0,1,1)(0,1,1)_{12}}_{}
    \;\xleftrightarrow{\text{Cleveland--Tiao (1976)}}\;
    \underbrace{\text{X-11 filter}}_{\text{current NSO practice}}
$}
\end{equation*}
This chain establishes BSM$(V_t=0)$ as a principled model-based analogue of
X-11-style seasonal adjustment: when sampling variances are negligible, the two
procedures deliver asymptotically identical point estimates for trend, seasonal,
and seasonally adjusted series.
It is important to note, however, that BSM$(V_t=0)$ is not literally X-11.
X-11 incorporates additional operational features---automatic outlier detection,
trading-day and Easter adjustment, endpoint revision protocols, and the full
battery of M-statistics and spectral diagnostics---that the BSM does not
replicate. The comparison is between DMM-BSM and X-11 as a seasonal
decomposition procedure, holding these additional features fixed.

Model-based methods for estimation from repeated surveys using state-space techniques were developed by \citet{pfeffermann1991}; composite estimation for rotating-panel surveys is studied by \citet{singh2001}, among others.

\citet{tam2026} introduced the DMM framework with an AR(1) propagation model,
which captures temporal persistence but cannot represent seasonal dynamics.
The present paper replaces the AR(1) with a BSM whose $12$-dimensional state
separates trend from seasonal components, enabling seasonal decomposition within
the Bayesian forward-filter framework. The three-model hierarchy (sampling, linking, and propagation
models) and the Sequential Hierarchical Bayesian Updating (SHBU) computational scheme are carried forward from
\citet{tam2026} with the propagation component replaced; all other
components are unchanged.

The central contribution of this paper is the combination of BSM propagation
with direct sampling-error incorporation via the time-varying measurement
variance $V_t^d$, embedded within a fully Bayesian Gibbs sampler.
A two-block Gibbs sampler alternates between (i) a joint state trajectory draw
via the Forward Filter Backward Sampler (\citealt{carterKohn1994};
\citealt{fruehwirthschnatter1994}), which delivers a draw from the exact joint
smoothing posterior $\pi(\bm{\alpha}_{1:T} \mid \mathbf{y}_{1:T})$ conditional
on the current hyperparameters, and (ii) conjugate inverse-gamma draws for the
variance hyperparameters $\sigma_\xi^2$ and $\sigma_\omega^2$ conditional on
the state trajectory from block (i).
This two-block scheme is a valid Gibbs sampler because the FFBS block (i)
draws the full state trajectory jointly, so the state differences
$L_t - L_{t-1}$ entering the hyperparameter full conditionals in block (ii)
have the correct joint posterior distribution.

This paper makes four contributions.

\begin{enumerate}
    \item \textbf{BSM propagation.} The AR(1) scalar state of \citet{tam2026}
    is replaced by the BSM $12$-dimensional state, separating trend and
    seasonal dynamics. The Gibbs full conditionals remain in closed conjugate
    form because the BSM system matrices $\mathbf{T}$ and $\mathbf{R}$ contain
    no free parameters (Section~\ref{sec:bsm}), so the sampler is pure
    Gibbs with no Metropolis steps.

    \item \textbf{Sampling error incorporation.} The SHBU posterior variance
    $\Sigma_t^{d,v}$ enters the Kalman filter measurement equation
    (equation~\eqref{eq:sampling_shbu}); the published sampling variance
    $V_t^{d,v}$ serves as the limiting case when stratum-level SHBU estimates
    are unavailable (equation~\eqref{eq:sampling_pervar}). The Kalman gain
    automatically down-weights high-CV months relative to the structural BSM
    prediction and gives near-full weight to low-CV months. This
    down-weighting mechanism is entirely absent from X-11.

    \item \textbf{FFBS Gibbs sampler with exact $k$-step movement inference.}
    The two-block Gibbs sampler uses the FFBS algorithm
    (Section~\ref{sec:gibbs}) to draw the full state trajectory jointly from
    the smoothing posterior, validating the hyperparameter full conditionals
    in Block~2. From the resulting smoothed trajectory draws
    $\{(L_1^{d,(b)}, \ldots, L_T^{d,(b)})\}_{b=1}^B$, the Calibrated Bayes
    Interval for $k$-step trend movement
    $\Delta_t^{d,(k)} = L_t^d - L_{t-k}^d$ and the posterior probability
    $P(\Delta_t^{d,(k)} > 0 \mid \mathbf{y}_{1:T})$ follow directly for any
    $k \geq 1$ as sample quantiles of the trajectory draws---no closed-form
    derivation required for any $k$, and all unavailable from X-11.

    \item \textbf{Verified coverage superiority.} A simulation study
    (Section~\ref{sec:coverage}) with $M=500$ replications and within-replication
    FFBS Gibbs parameter estimation demonstrates substantially higher empirical
    coverage than the X-11-equivalent across both domains.
    For the large domain (Australia, mean CV 0.38\%), DMM-BSM achieves
    94.6\%/94.5\% trend-level/change coverage; for the small domain
    (ACT, mean CV 1.62\%), 86.7\%/82.1\%---a setting where standard
    large-sample approximations can fail \citep{rao2015}.
    The X-11-equivalent model that ignores sampling error achieves only
    78.8\%/64.3\% and 68.2\%/50.8\%, respectively, with the shortfall largest
    for the small domain.
\end{enumerate}

The empirical illustration uses 120 months (January 2014--December 2023) of
full-time employment estimates from the ABS Labour Force Survey, with $V = 1$
variable (full-time employed persons) and $D = 2$ domains: Australia (AU) as
the large domain and the Australian Capital Territory (ACT) as the small
domain. Domain-level standard errors are available for both series and are
supplied directly to the Kalman filter. The stratum-level data required for
the full DMM mini-max sample allocation of \citet{tam2026} are not available
for this illustration; the empirical section therefore focuses on the
propagation and uncertainty-quantification components of the framework, which
operate at the domain level.

\section{The DMM-BSM Framework}
\label{sec:model}

Let $d = 1, \ldots, D$ index strata,
$t = 1, \ldots, T$ index survey waves,
and $v = 1, \ldots, V$ index survey variables.
In the empirical illustration of Section~\ref{sec:empirical}, $d$ indexes
the two domains (ACT and Australia) directly; stratum-level data are not
available for that illustration. At each wave $t$ and domain $d$,
the survey produces a $V$-vector of direct estimates
\begin{equation}
    \mathbf{Y}_t^d = (Y_t^{d,1}, \ldots, Y_t^{d,V})^\top \in \mathbb{R}^V,
    \label{eq:obs_vector}
\end{equation}
with an associated $V \times V$ design-based sampling covariance matrix
$\mathbf{V}_t^d$, treated as known. The diagonal entries
$V_t^{d,vv} = (\mathrm{SE}_t^{d,v})^2$ are the squared standard errors for
each variable; off-diagonal entries capture within-stratum cross-variable
sampling covariance.

The DMM-BSM comprises the same three-model hierarchy as \citet{tam2026}:
a sampling model, a linking model, and a propagation model. The first two
components are unchanged; only the propagation model is replaced.

\subsection{Sampling Model}
\label{sec:sampling}

For each variable $v$, the sampling model relates the direct survey estimate to
the latent BSM state:
\begin{equation}
    Y_t^{d,v} = \mathbf{Z}^\top \bm{\alpha}_t^{d,v} + e_t^{d,v}, \qquad
    e_t^{d,v} \sim \mathcal{N}(0,\, V_t^{d,v}), \quad V_t^{d,v} \text{ known},
    \label{eq:sampling_pervar}
\end{equation}
where $\bm{\alpha}_t^{d,v} \in \mathbb{R}^{12}$ is the BSM state vector for
variable $v$ defined in Section~\ref{sec:bsm}, $\mathbf{Z}$ is the selection
vector, and $V_t^{d,v} = (\mathrm{SE}_t^{d,v})^2$ is the known sampling
variance. In stacked form across variables:
\begin{equation}
    \mathbf{Y}_t^d = \tilde{\mathbf{Z}}\, \tilde{\bm{\alpha}}_t^d
        + \mathbf{e}_t^d, \qquad
    \mathbf{e}_t^d \sim \mathcal{N}(\mathbf{0},\, \mathbf{V}_t^d),
    \label{eq:sampling_vector}
\end{equation}
where $\tilde{\bm{\alpha}}_t^d = (\bm{\alpha}_t^{d,1\top}, \ldots,
\bm{\alpha}_t^{d,V\top})^\top \in \mathbb{R}^{12V}$ and
$\tilde{\mathbf{Z}} = \mathbf{I}_V \otimes \mathbf{Z}^\top$ is the
$V \times 12V$ block-diagonal selection matrix. Setting
$\mathbf{V}_t^d \equiv \mathbf{0}$ recovers X-11's implicit assumption that
observations are exact.

\begin{remark}
For $V = 1$, equation~\eqref{eq:sampling_vector} reduces to the scalar case
$Y_t^d = \mathbf{Z}^\top \bm{\alpha}_t^d + e_t^d$ with
$V_t^d = (\mathrm{SE}_t^d)^2$. This is the case used in the empirical
illustration of Section~\ref{sec:empirical}.
The general $V > 1$ formulation, including off-diagonal cross-variable
sampling covariances in $\mathbf{V}_t^d$ (same respondent answering
multiple questions in the same wave), is stated for generality;
joint inference across multiple variables exploiting the full $\mathbf{V}_t^d$
matrix is not implemented in this paper and is reserved for future work.
\end{remark}

\begin{remark}[Serially correlated sampling errors]
\label{rem:sampling_covariance}
The sampling model \eqref{eq:sampling_pervar} treats measurement errors
$e_t^{d,v}$ as independent across waves.
For rotating-panel labour force surveys such as the ABS LFS, adjacent-wave
sampling errors are positively correlated through sample overlap: a portion
of respondents appears in both wave $t$ and wave $t-1$, inducing non-zero
$\mathrm{Cov}(e_t^{d,v}, e_{t-1}^{d,v})$.
The design-based variance of the wave-to-wave change is therefore
$\mathrm{Var}(Y_t^{d,v} - Y_{t-1}^{d,v})
 = V_t^{d,v} + V_{t-1}^{d,v} - 2\,\mathrm{Cov}(e_t^{d,v}, e_{t-1}^{d,v})$,
and ignoring the positive cross-term overstates movement uncertainty, yielding
movement CBIs that are conservative (wider than the nominal 95\% target).
In the present paper, this covariance is set to zero because the ABS does
not publicly release inter-wave sampling covariance matrices;
the following analysis quantifies the magnitude of this conservatism.
Estimation of inter-wave sampling covariances for rotating panels is studied by \citet{pfeffermann1998}.

Under the ABS LFS 8-rotation-group design, the sample overlap between waves
$t$ and $t-k$ is $(8-k)/8$ for $k = 1,\ldots,7$ and zero for $k \geq 8$.
By rotating-panel sampling theory \citep{pfeffermann1998}, the inter-wave
sampling covariance for approximately constant sampling variance $V$ is
\begin{equation*}
    \mathrm{Cov}(e_t^{d,v}, e_{t-k}^{d,v}) \approx
    \frac{8-k}{8}\,\rho^{(v)}\,V, \quad k = 1,\ldots,7,
\end{equation*}
where $\rho^{(v)} \in [0,1]$ is the between-wave correlation of the
characteristic for matched respondents.
\citet{tam2026} calibrated $\rho^{(v)}$ to 2021 Australian Census microdata
for employment and unemployment, obtaining values in the range
$\rho^{(v)} \approx 0.4$--$0.6$.
The corrected wave-to-wave ($k=1$) movement variance is
$2V\bigl(1 - \tfrac{7}{8}\rho^{(v)}\bigr)$, compared to $2V$ under independence;
the ratio $\bigl[1 - \tfrac{7}{8}\rho^{(v)}\bigr]^{-1}$
evaluates to $1.54$--$2.11$ over $\rho^{(v)} \in [0.4,\,0.6]$,
implying the independence assumption overstates the wave-to-wave movement
standard deviation by 24--45\%.
For quarterly movement ($k=3$) the overlap falls to $5/8$, giving an
overstatement of 15--26\%.
For $k \geq 8$ no respondents overlap and independence holds exactly.
In all cases the bias is conservative: ignoring the positive covariance
widens the movement CBI beyond its nominal 95\% target, so published
intervals remain valid but are unnecessarily wide.
\end{remark}

\subsection{Linking Model}
\label{sec:linking}

The linking model is inherited unchanged from \citet{tam2026}; the specification
below is restated here for completeness and to define the full DMM-BSM hyperparameter
set $\bm{\psi}$. The level component $L_t^{d,v}$ of the state
vector satisfies a hierarchical regression that borrows strength across strata:
\begin{equation}
    L_t^{d,v} = \mathbf{x}_t^{d\top} \bm{\beta}^v + u_t^{d,v}, \qquad
    u_t^{d,v} \sim \mathcal{N}(0,\, \sigma_{u,v}^2), \quad v = 1,\ldots,V,
    \label{eq:linking}
\end{equation}
where $\mathbf{x}_t^d$ is a vector of auxiliary variables for stratum $d$
at wave $t$, $\bm{\beta}^v$ is a regression coefficient vector shared across
all strata for variable $v$, and $\sigma_{u,v}^2$ is the between-stratum
linking variance for variable $v$.

The linking model borrows strength across strata through the shared
$\bm{\beta}^v$: the minimum sample size reduction across all target variables
that simultaneously satisfies level and movement precision constraints defines
the DMM mini-max allocation of \citet{tam2026}.
The empirical illustration of Section~\ref{sec:empirical} does not use
stratum-level data, so the two domains are fitted independently using
domain-level direct estimates; the linking model and mini-max allocation
are available for future application when stratum-level SHBU outputs are available.

At each wave $t$, the Sequential Hierarchical Bayesian Updating (SHBU) of
\citet{tam2026} combines the stratum-level direct estimates
$\{Y_t^{h,v}\}_{h \in \mathcal{H}_d}$ with the linking model
\eqref{eq:linking} to produce a stratum-level posterior mean
\begin{equation}
    \hat{Y}_t^{d,v,\mathrm{HB}}
    = \frac{\displaystyle\sum_{h \in \mathcal{H}_d}
      N_h\,\hat{\theta}_{h,t}^{v,\mathrm{HB}}}{N_d}
    \label{eq:yhat_link}
\end{equation}
and posterior variance $\Sigma_t^{d,v}$, where
$\hat{\theta}_{h,t}^{v,\mathrm{HB}}$ is the SHBU stratum-level posterior
mean.  The BSM measurement equation \eqref{eq:sampling_pervar} then takes
$\hat{Y}_t^{d,v,\mathrm{HB}}$ as the observation:
\begin{equation}
    \hat{Y}_t^{d,v,\mathrm{HB}} = \mathbf{Z}^\top \bm{\alpha}_t^{d,v}
    + \tilde{e}_t^{d,v}, \qquad
    \tilde{e}_t^{d,v} \sim \mathcal{N}(0,\, \Sigma_t^{d,v}),
    \label{eq:sampling_shbu}
\end{equation}
with $\hat{Y}_t^{d,v,\mathrm{HB}}$ in place of $Y_t^{d,v}$ and
$\Sigma_t^{d,v}$ in place of $V_t^{d,v}$.
The BSM state dynamics \eqref{eq:level}--\eqref{eq:seasonal} are unchanged;
all Kalman filter and FFBS equations carry through with $\Sigma_t^{d,v}$
substituted for $V_t^{d,v}$.

In the absence of SHBU outputs --- for example, when only the published
aggregate direct estimate is available --- the formulation reduces to the
standard sampling model \eqref{eq:sampling_pervar} by setting
$\hat{Y}_t^{d,v,\mathrm{HB}} = Y_t^{d,v}$ and
$\Sigma_t^{d,v} = V_t^{d,v}$.

\begin{remark}[Estimated variances treated as known]
\label{rem:known_variance}
Both formulations treat the measurement error variance as a known fixed
quantity in the Kalman filter: $\Sigma_t^{d,v}$ in the SHBU pipeline and
$V_t^{d,v}$ in the limiting case. In practice both are estimated from the
survey data and carry their own uncertainty. Treating estimated variances as
known in the measurement equation is, however, the standard approximation
throughout the survey-based Kalman filter literature: the alternative of
modelling variance estimation uncertainty jointly with the state dynamics
greatly increases model complexity for limited practical gain when sample
sizes are moderate to large. The SHBU pipeline introduces no additional
approximation relative to the direct-survey case in this respect.
\end{remark}

\subsection{BSM Propagation Model}
\label{sec:bsm}

In \citet{tam2026} the propagation model is the AR(1):
\begin{equation}
    \theta_t^d = \rho \theta_{t-1}^d + (1-\rho)\mu + \eta_t^d, \qquad
    \eta_t^d \sim \mathcal{N}(0,\, \sigma_\eta^2).
    \label{eq:ar1}
\end{equation}
This is adequate for modelling smooth temporal persistence but cannot represent
seasonal dynamics. For monthly employment series where seasonal variation
dominates short-run fluctuations, a model that separates trend from seasonal
components is required.

We replace \eqref{eq:ar1} with a Basic Structural Model comprising a
local level trend and a stochastic seasonal component.
For monthly data ($s = 12$), the state vector for variable $v$, domain $d$ is
\begin{equation}
    \bm{\alpha}_t^{d,v} =
    \bigl(L_t^{d,v},\; \gamma_t^{d,v},\; \gamma_{t-1}^{d,v},\;
    \ldots,\; \gamma_{t-10}^{d,v}\bigr)^\top \in \mathbb{R}^{12},
    \label{eq:state_vector}
\end{equation}
where $L_t^{d,v}$ is the unobserved trend level and $\gamma_t^{d,v}$ is the
current seasonal effect. The state is $12$-dimensional because the stochastic
seasonal constraint \eqref{eq:seasonal} involves all $s-1 = 11$ previous
seasonal values; stacking these lags into the state vector is the standard
device for expressing the constraint as a first-order Markov system. The BSM propagation equations are:
\begin{align}
    L_t^{d,v} &= L_{t-1}^{d,v} + \xi_t^{d,v}, \quad
    \xi_t^{d,v} \sim \mathcal{N}(0,\,\sigma_{\xi,v}^2),
    \label{eq:level} \\
    \gamma_t^{d,v} &= -\sum_{j=1}^{s-1} \gamma_{t-j}^{d,v} + \omega_t^{d,v},
    \quad \omega_t^{d,v} \sim \mathcal{N}(0,\,\sigma_{\omega,v}^2),
    \label{eq:seasonal}
\end{align}
with all disturbances mutually independent across variables, strata, and waves.
In matrix form:
\begin{equation}
    \bm{\alpha}_t^{d,v} = \mathbf{T}\, \bm{\alpha}_{t-1}^{d,v}
        + \mathbf{R}\, \bm{\eta}_t^{d,v}, \quad
    \bm{\eta}_t^{d,v} \sim \mathcal{N}(\mathbf{0},\, \mathbf{Q}_v),
    \label{eq:state_eq}
\end{equation}
where the system matrices are:
\begin{equation}
    \mathbf{T} = \begin{pmatrix}
        1 & 0 & 0 & \cdots & 0 & 0 \\
        0 & -1 & -1 & \cdots & -1 & -1 \\
        0 & 1 & 0 & \cdots & 0 & 0 \\
        0 & 0 & 1 & \cdots & 0 & 0 \\
        \vdots & & & \ddots & & \vdots \\
        0 & 0 & 0 & \cdots & 1 & 0
    \end{pmatrix} \in \mathbb{R}^{12\times 12}, \qquad
    \mathbf{Z} = \begin{pmatrix}1\\1\\0\\\vdots\\0\end{pmatrix}
    \in \mathbb{R}^{12},
    \label{eq:T_matrix}
\end{equation}
The observation selection vector $\mathbf{Z}$ selects the sum of the level
and current seasonal component from the state; it appears in
equations~\eqref{eq:sampling_pervar}, \eqref{eq:innovation}--\eqref{eq:gain}.
\begin{equation}
    \mathbf{R} = \begin{pmatrix} 1 & 0 \\ 0 & 1 \\ 0 & 0 \\ \vdots & \vdots \\
    0 & 0 \end{pmatrix} \in \mathbb{R}^{12\times 2}, \qquad
    \mathbf{Q}_v = \begin{pmatrix} \sigma_{\xi,v}^2 & 0 \\
    0 & \sigma_{\omega,v}^2 \end{pmatrix} \in \mathbb{R}^{2\times 2}.
    \label{eq:RQ}
\end{equation}
The hyperparameters to be estimated are
$\bm{\psi} = \{\sigma_{\xi,v}^2,\, \sigma_{\omega,v}^2,\,
\sigma_{u,v}^2,\, \bm{\beta}^v\}_{v=1}^V$.

\section{Equivalence with X-11}
\label{sec:equivalence}

\citet{harveytod1983} showed that the reduced form of the BSM with local
level and stochastic seasonal, obtained by applying the differencing operator
$(1-B)(1-B^{12})$ to \eqref{eq:sampling_vector} with
$\mathbf{V}_t^d \equiv \mathbf{0}$, is the airline model
\begin{equation}
    (1-B)(1-B^{12})\, Y_t^d =
    (1 - \theta B)(1 - \Theta B^{12})\, \varepsilon_t, \quad
    \varepsilon_t \sim \mathcal{N}(0,\, \sigma_\varepsilon^2),
    \label{eq:airline}
\end{equation}
where the MA parameters $(\theta, \Theta)$ are one-to-one functions of the
BSM signal-to-noise ratios $q_\xi = \sigma_\xi^2 / \sigma_\varepsilon^2$
and $q_\omega = \sigma_\omega^2 / \sigma_\varepsilon^2$.\footnote{The BSM
used in this paper omits the observation irregular
$\varepsilon_t \sim \mathcal{N}(0,\,\sigma_\varepsilon^2)$ present in
\citet[][Ch.~2]{harvey1989}: observation noise enters only through the
survey sampling variance $V_t^d$. When $V_t^d \equiv 0$, the observation
irregular is absent and the reduced form is the constrained airline model
with $\sigma_\varepsilon^2 = 0$; the MA parameters $(\theta,\Theta)$ are
then determined solely by $(\sigma_\xi^2,\sigma_\omega^2)$.
The Harvey--Todd equivalence applies to this constrained BSM.}
Combined with \citet{clevelandtiao1976}, the equivalence chain gives:
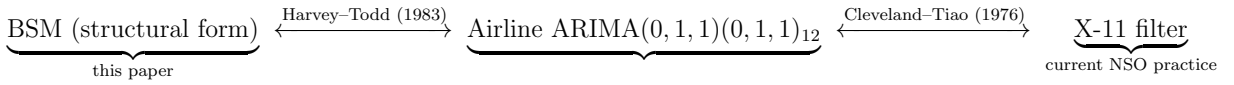
\begin{equation*}
\resizebox{\textwidth}{!}{$
    \underbrace{\text{BSM (structural form)}}_{\text{this paper}}
    \;\xleftrightarrow{\text{Harvey--Todd (1983)}}\;
    \underbrace{\text{Airline ARIMA}(0,1,1)(0,1,1)_{12}}_{}
    \;\xleftrightarrow{\text{Cleveland--Tiao (1976)}}\;
    \underbrace{\text{X-11 filter}}_{\text{current NSO practice}}
$}
\end{equation*}

This equivalence anchors the comparison between DMM-BSM and X-11.
When sampling variances are negligible ($V_t^d \approx 0$), the corresponding
BSM and X-11 produce asymptotically identical trend, seasonal, and seasonally
adjusted estimates. The corresponding BSM is therefore a principled model-based
benchmark for assessing the additional effect of incorporating survey sampling
error: any difference between DMM-BSM and BSM$(V_t=0)$ is attributable solely
to the $\Sigma_t^{d,v}$ term in the measurement equation, not to a difference
in the underlying decomposition model.

It bears emphasis that BSM$(V_t=0)$ is a structural-model analogue of
X-11, not a literal reimplementation. X-11 encompasses additional operational
capabilities---automatic outlier and calendar-effect adjustment, a battery of
diagnostics (M-statistics, spectral checks), and endpoint revision
protocols---that are outside the BSM framework.

The equivalence also defines the empirical comparison strategy.
Since X-11 produces no uncertainty measures for trend or trend movement,
any gain from DMM-BSM is entirely in uncertainty quantification rather
than point estimation. In the empirical illustration
(Section~\ref{sec:empirical}), BSM$(V_t = 0)$ serves as the
X-11-equivalent benchmark; the comparison with DMM-BSM
(measurement error variance $\Sigma_t^{d,v}$) isolates the effect of
incorporating survey sampling error directly.

\section{Sequential HB Updating as Bayesian Forward Filtering}
\label{sec:shbu}

\subsection{SHBU-BSM as Bayesian Forward Filter}
\label{sec:ff}

At each wave $t$, the SHBU performs two steps.

\textbf{Propagation step.} Let
$\bar{\bm{\alpha}}_{t-1} = \E[\bm{\alpha}_{t-1} \mid \mathbf{y}_{1:t-1}]
= \bm{\alpha}_{t-1|t-1}$
denote the wave $t-1$ filtered posterior mean.
The predictive distribution at wave $t$ is obtained by integrating the BSM
transition over the wave $t-1$ filtered posterior:
\begin{equation}
    \pi(\bm{\alpha}_t \mid \mathbf{y}_{1:t-1})
    \;=\; \int p(\bm{\alpha}_t \mid \bm{\alpha}_{t-1})\,
    \pi(\bm{\alpha}_{t-1} \mid \mathbf{y}_{1:t-1})\,
    d\bm{\alpha}_{t-1}.
    \label{eq:predict}
\end{equation}
For the linear Gaussian BSM, this integral is exact and evaluates to
\begin{equation*}
    \pi(\bm{\alpha}_t \mid \mathbf{y}_{1:t-1})
    = \mathcal{N}\!\left(\mathbf{T}\bm{\alpha}_{t-1|t-1},\;
    \mathbf{T}\mathbf{P}_{t-1|t-1}\mathbf{T}^\top
    + \mathbf{R}\mathbf{Q}\mathbf{R}^\top\right),
\end{equation*}
which is precisely the Kalman prediction step
\eqref{eq:predict_state}--\eqref{eq:predict_cov}.
The full posterior covariance $\mathbf{P}_{t-1|t-1}$ is propagated forward
through $\mathbf{T}\mathbf{P}_{t-1|t-1}\mathbf{T}^\top$; no between-draw
variance is omitted.
This propagated distribution serves as the prior at wave $t$.

\textbf{Update step.} The wave $t$ observation $\mathbf{y}_t$ updates the prior
via Bayes' rule:
\begin{equation}
    \pi(\bm{\alpha}_t \mid \mathbf{y}_{1:t}) \propto
    p(\mathbf{y}_t \mid \bm{\alpha}_t)\,
    \pi(\bm{\alpha}_t \mid \mathbf{y}_{1:t-1}),
    \label{eq:update}
\end{equation}
For the linear Gaussian BSM, with Gaussian prior
$\pi(\bm{\alpha}_t \mid \mathbf{y}_{1:t-1})
= \mathcal{N}(\bm{\alpha}_{t|t-1}^{d,v},\,\mathbf{P}_{t|t-1}^{d,v})$
and Gaussian likelihood
$p(\mathbf{y}_t \mid \bm{\alpha}_t)
= \mathcal{N}(\mathbf{Z}^\top\bm{\alpha}_t,\,\Sigma_t^{d,v})$,
completing the square in the exponent yields the Gaussian filtered posterior
\begin{equation*}
    \pi(\bm{\alpha}_t \mid \mathbf{y}_{1:t})
    = \mathcal{N}\!\bigl(\bm{\alpha}_{t|t}^{d,v},\,\mathbf{P}_{t|t}^{d,v}\bigr),
\end{equation*}
with $\bm{\alpha}_{t|t}^{d,v}$ and $\mathbf{P}_{t|t}^{d,v}$ given by
equations~\eqref{eq:update_state}--\eqref{eq:update_cov} in
Section~\ref{sec:kalman}. Equation~\eqref{eq:update} delivers the
filtered marginal $\pi(\bm{\alpha}_t \mid \mathbf{y}_{1:t})$, not the
Gibbs full conditional, which is the joint smoothing posterior
$\pi(\bm{\alpha}_{1:T} \mid \bm{\psi},\,\mathbf{y}_{1:T})$
sampled by FFBS in Block~1 (Section~\ref{sec:gibbs}).
The filtered marginal is the key input to the FFBS backward conditional.

These two steps are repeated for $t = 1, 2, \ldots, T$, conditioning only on
data available up to wave $t$. The process is purely forward in time and
describes the conceptual structure of sequential Bayesian updating.

\subsection{Kalman Filter Prediction and Update}
\label{sec:kalman}

Conditional on the hyperparameters $\bm{\psi}$, the propagation step
\eqref{eq:predict} and update step \eqref{eq:update} are implemented via the
Kalman filter \citep{durbin2012,commandeur2007}. For each domain $d$ and variable $v$,
the forward pass at wave $t$ proceeds as follows.

\textbf{Prediction:}
\begin{align}
    \bm{\alpha}_{t|t-1}^{d,v} &= \mathbf{T}\, \bm{\alpha}_{t-1|t-1}^{d,v},
    \label{eq:predict_state} \\
    \mathbf{P}_{t|t-1}^{d,v} &= \mathbf{T}\, \mathbf{P}_{t-1|t-1}^{d,v}\,
        \mathbf{T}^\top + \mathbf{R}\,\mathbf{Q}_v\,\mathbf{R}^\top.
    \label{eq:predict_cov}
\end{align}

\textbf{Update:}
\begin{align}
    \nu_t^{d,v} &= Y_t^{d,v} - \mathbf{Z}^\top \bm{\alpha}_{t|t-1}^{d,v},
    \label{eq:innovation} \\
    F_t^{d,v} &= \mathbf{Z}^\top \mathbf{P}_{t|t-1}^{d,v} \mathbf{Z}
        + V_t^{d,v},
    \label{eq:innov_var} \\
    \mathbf{K}_t^{d,v} &= \mathbf{P}_{t|t-1}^{d,v}\, \mathbf{Z}\,
        (F_t^{d,v})^{-1},
    \label{eq:gain} \\
    \bm{\alpha}_{t|t}^{d,v} &= \bm{\alpha}_{t|t-1}^{d,v}
        + \mathbf{K}_t^{d,v}\, \nu_t^{d,v},
    \label{eq:update_state} \\
    \mathbf{P}_{t|t}^{d,v} &= (\mathbf{I} - \mathbf{K}_t^{d,v}\,
        \mathbf{Z}^\top)\, \mathbf{P}_{t|t-1}^{d,v}.
    \label{eq:update_cov}
\end{align}

The Kalman gain $\mathbf{K}_t^{d,v}$ in \eqref{eq:gain} makes the role of
sampling error explicit: when $V_t^{d,v}$ is large (high-CV wave), $F_t^{d,v}$
increases, the gain decreases, and the filter relies more heavily on the
structural BSM prediction than on the observed survey estimate. When
$V_t^{d,v} \to 0$, the filter reduces to the X-11 signal extraction formula
(by the Harvey--Todd equivalence). The difference between the two cases is
precisely the effect of sampling error on seasonal adjustment---an effect that
X-11 cannot represent.

\subsection{Backward Kalman Step for Joint Movement Draws}
\label{sec:backward}

The forward Kalman pass yields the filtered marginal posterior at each
wave $t$: $\pi(\bm{\alpha}_t \mid \mathbf{y}_{1:t})
= \mathcal{N}(\bm{\alpha}_{t|t},\,\mathbf{P}_{t|t})$.
For movement inference the quantity of interest is the joint filtered
posterior of the trend at two consecutive waves,
$\pi(L_{t-1}^{d,v},\,L_t^{d,v} \mid \mathbf{y}_{1:t})$.
This joint distribution cannot be recovered by pairing independent draws from
the wave-$t{-}1$ and wave-$t$ chains: the wave-$t{-}1$ chain was run without
knowledge of $\mathbf{y}_t$, so its marginal is
$\pi(\bm{\alpha}_{t-1}\mid\mathbf{y}_{1:t-1})$, not the posterior conditional
on the full current data $\mathbf{y}_{1:t}$.
The joint filtered posterior is instead obtained via a single backward
Kalman step applied to the immediately preceding state, using only quantities
already computed in the forward pass.

Given the Kalman forward-pass quantities at waves $t{-}1$ and $t$, define the
backward smoother gain
\begin{equation}
    \mathbf{J}_{t-1}^{d,v}
    \;=\; \mathbf{P}_{t-1|t-1}^{d,v}\,\mathbf{T}^\top
          \bigl(\mathbf{P}_{t|t-1}^{d,v}\bigr)^{-1}.
    \label{eq:backward_gain}
\end{equation}
The one-step backward conditional for the full state vector is then
\begin{equation}
    \pi\!\left(\bm{\alpha}_{t-1} \mid \bm{\alpha}_t,\,
               \mathbf{y}_{1:t-1}\right)
    = \mathcal{N}\!\left(
        \mathbf{m}_{t-1|t}^{d,v}(\bm{\alpha}_t),\;
        \mathbf{C}_{t-1|t}^{d,v}
      \right),
    \label{eq:backward_cond}
\end{equation}
where
\begin{align}
    \mathbf{m}_{t-1|t}^{d,v}(\bm{\alpha}_t)
        &= \bm{\alpha}_{t-1|t-1}^{d,v}
         + \mathbf{J}_{t-1}^{d,v}
           \bigl(\bm{\alpha}_t - \bm{\alpha}_{t|t-1}^{d,v}\bigr),
    \label{eq:backward_mean} \\
    \mathbf{C}_{t-1|t}^{d,v}
        &= \mathbf{P}_{t-1|t-1}^{d,v}
         - \mathbf{J}_{t-1}^{d,v}\,\mathbf{P}_{t|t-1}^{d,v}\,
           \bigl(\mathbf{J}_{t-1}^{d,v}\bigr)^\top.
    \label{eq:backward_var}
\end{align}
All quantities in \eqref{eq:backward_gain}--\eqref{eq:backward_var} are
available directly from the Kalman forward pass
\eqref{eq:predict_state}--\eqref{eq:update_cov}; the backward step requires
only a single additional matrix multiplication and draw per posterior sample,
using no data beyond wave $t$.

To obtain exact paired draws from
$\pi(L_{t-1}^{d,v},\,L_t^{d,v} \mid \mathbf{y}_{1:t})$, one proceeds as
follows for each posterior draw $b = 1, \ldots, B$ from the wave-$t$ Gibbs
chain:
\begin{enumerate}[label=(\roman*)]
    \item Take $\bm{\alpha}_t^{d,v,(b)}$ as the wave-$t$ state vector
          already produced by the Gibbs Block~1 FFBS pass
          (Section~\ref{sec:gibbs}) at the current MCMC iteration.
    \item Draw $\bm{\alpha}_{t-1}^{d,v,(b)\star}
          \sim \mathcal{N}\!\bigl(
              \mathbf{m}_{t-1|t}^{d,v}(\bm{\alpha}_t^{d,v,(b)}),\;
              \mathbf{C}_{t-1|t}^{d,v}
          \bigr)$
          from \eqref{eq:backward_cond}.
    \item Set $L_{t-1}^{d,v,(b)\star}
          = \mathbf{e}_1^\top\bm{\alpha}_{t-1}^{d,v,(b)\star}$,
          where $\mathbf{e}_1 = (1,0,\ldots,0)^\top \in \mathbb{R}^m$
          selects the trend level from the state vector (see also
          \eqref{eq:extract_components}), and
          $\Delta_t^{d,v,(b)}
          = L_t^{d,v,(b)} - L_{t-1}^{d,v,(b)\star}$.
\end{enumerate}
The pairs
$\bigl\{(L_{t-1}^{d,v,(b)\star},\,L_t^{d,v,(b)})\bigr\}_{b=1}^B$
are exact Monte Carlo draws, under the fitted linear Gaussian BSM, from
the one-step joint filtered posterior
$\pi(L_{t-1}^{d,v},\,L_t^{d,v} \mid \mathbf{y}_{1:t})$.
The naive alternative of pairing draws
$(L_{t-1}^{d,v,(b)},\,L_t^{d,v,(b)})$ from two independent Gibbs chains
yields instead the product marginal
$\pi(L_{t-1}^{d,v}\mid\mathbf{y}_{1:t-1})\,
 \pi(L_t^{d,v}\mid\mathbf{y}_{1:t})$,
which ignores the posterior covariance induced by the shared BSM transition
and therefore overstates $\mathrm{Var}(\Delta_t^{d,v}\mid\mathbf{y}_{1:t})$.

The equal-tailed credible interval for the $k$-step movement is
\begin{equation}
    \mathrm{CBI}_{t,k}^{d,v}
    = \bigl[Q_{0.025}(\Delta_t^{d,v,(b)}),\;
             Q_{0.975}(\Delta_t^{d,v,(b)})\bigr],
    \label{eq:cbi}
\end{equation}
where $Q_p$ denotes the $p$th sample quantile over the $B$ Gibbs draws,
and $\Delta_t^{d,v,(b)} = L_t^{d,v,(b)} - L_{t-1}^{d,v,(b)\star}$
from the paired backward-step draws.

\subsection{Gibbs Sampler: State Draws and Full Conditionals}
\label{sec:gibbs}

DMM-BSM requires the full joint posterior trajectory
$\pi(\bm{\alpha}_{1:T} \mid \bm{\psi}, \mathbf{y}_{1:T})$, not merely
the wave-by-wave filtered marginals. The forward Kalman pass
(Section~\ref{sec:kalman}) yields, at each wave $t$, only the filtered
marginal $\pi(\bm{\alpha}_t \mid \mathbf{y}_{1:t})$.
Two requirements make the full trajectory indispensable.
First, the Block~2 hyperparameter updates for $\sigma_\xi^2$ and
$\sigma_\omega^2$ depend on the level differences $L_t - L_{t-1}$
accumulated over all $T$ waves: computing their distribution requires the
joint smoothing posterior $\pi(\bm{\alpha}_{1:T} \mid \bm{\psi},
\mathbf{y}_{1:T})$, not a product of independent filtered marginals,
which would assign zero cross-time covariance and distort the Block~2 update.
Second, $k$-step trend movement CBIs, sustained-trend probabilities,
threshold crossings, and cumulative change distributions are direct
by-products of the joint posterior trajectory, requiring no additional
derivation once the full draw $\{L_t^{d,(b)}\}_{t=1}^T$ is in hand.

The Forward Filter Backward Sampler \citep{carterKohn1994,
fruehwirthschnatter1994} (FFBS) provides exact draws from this joint
posterior in two passes: a forward Kalman filter pass storing the filtered
quantities at each $t$, followed by a backward sampling pass that
sequentially draws each $\bm{\alpha}_t$ conditional on $\bm{\alpha}_{t+1}$
already drawn.
The one-step backward conditional of Section~\ref{sec:backward} is a
single step of this backward pass, used for real-time ($k=1$) movement
inference without a full backward sweep; FFBS generalises it to the
complete trajectory, enabling all $k \geq 1$ and supporting the Block~2
update.
When a new wave $T+1$ arrives, FFBS is re-run over the extended series,
updating all smoothed estimates retrospectively.

The Gibbs sampler for DMM-BSM alternates between two blocks at each
MCMC iteration $k$.

\textbf{Step~1: FFBS state trajectory draw.}
Given the current hyperparameters
$\bm{\psi}^{(k)} = (\sigma_\xi^{2,(k)},\,\sigma_\omega^{2,(k)},\,
\sigma_u^{2,(k)},\,\bm{\beta}^{(k)})$,
draw the full state trajectory jointly from the smoothing posterior
$\pi(\bm{\alpha}_{1:T}^{d,v} \mid \mathbf{y}_{1:T}^d,\,\bm{\psi}^{(k)})$
via the Forward Filter Backward Sampler of \citet{carterKohn1994}.

(a) Kalman forward pass.
Run the Kalman filter (Section~\ref{sec:kalman}) forward from $t=1$ to $T$,
storing $\{\bm{\alpha}_{t|t}^{d,v},\,\mathbf{P}_{t|t}^{d,v},\,
\bm{\alpha}_{t+1|t}^{d,v},\,\mathbf{P}_{t+1|t}^{d,v}\}_{t=1}^{T}$.

(b) FFBS backward sampling pass.
Draw the terminal state:
\begin{equation}
    \bm{\alpha}_T^{d,v,(k)} \sim
    \mathcal{N}\!\left(\bm{\alpha}_{T|T}^{d,v},\;
    \mathbf{P}_{T|T}^{d,v}\right).
    \label{eq:ffbs_terminal}
\end{equation}
Then for $t = T{-}1, T{-}2, \ldots, 1$, compute the FFBS smoother gain
\begin{equation}
    \mathbf{J}_t^{d,v} = \mathbf{P}_{t|t}^{d,v}\,\mathbf{T}^\top
    \bigl(\mathbf{P}_{t+1|t}^{d,v}\bigr)^{-1},
    \label{eq:ffbs_gain}
\end{equation}
and draw
\begin{equation}
    \bm{\alpha}_t^{d,v,(k)} \sim \mathcal{N}\!\Bigl(
        \bm{\alpha}_{t|t}^{d,v}
        + \mathbf{J}_t^{d,v}\bigl(\bm{\alpha}_{t+1}^{d,v,(k)}
          - \bm{\alpha}_{t+1|t}^{d,v}\bigr),\;
        \mathbf{P}_{t|t}^{d,v}
        - \mathbf{J}_t^{d,v}\,\mathbf{P}_{t+1|t}^{d,v}\,
          \bigl(\mathbf{J}_t^{d,v}\bigr)^\top
    \Bigr).
    \label{eq:ffbs_draw}
\end{equation}
The resulting draw
$\bm{\alpha}_{1:T}^{d,v,(k)}
 = \bigl(\bm{\alpha}_1^{d,v,(k)},\ldots,\bm{\alpha}_T^{d,v,(k)}\bigr)$
is an exact Monte Carlo draw from the joint smoothing posterior
$\pi(\bm{\alpha}_{1:T}^{d,v} \mid \mathbf{y}_{1:T}^d,\,\bm{\psi}^{(k)})$
\citep{carterKohn1994, fruehwirthschnatter1994}.
Note that \eqref{eq:ffbs_gain}--\eqref{eq:ffbs_draw} use the same
smoother gain structure as the one-step backward conditional of
Section~\ref{sec:backward}, generalised to a full backward pass over all $T$
time points rather than a single step.

Extract the level and seasonal components as the first two elements of each
state draw:
\begin{equation}
    L_t^{d,v,(k)} = \mathbf{e}_1^\top\bm{\alpha}_t^{d,v,(k)},
    \qquad
    \gamma_t^{d,v,(k)} = \mathbf{e}_2^\top\bm{\alpha}_t^{d,v,(k)},
    \label{eq:extract_components}
\end{equation}
where $\mathbf{e}_1 = (1,0,\ldots,0)^\top$ and
$\mathbf{e}_2 = (0,1,0,\ldots,0)^\top$ are the first and second standard
basis vectors in $\mathbb{R}^{12}$.

\textbf{Step~2: Hyperparameter full conditionals.}
Conditional on the state draws $\{L_t^{d,v,(k)}, \gamma_t^{d,v,(k)}\}$
from Step~1, the hyperparameter full conditionals are in
closed form for each variable $v = 1, \ldots, V$:
\begin{align}
    \sigma_{\xi,v}^2 \mid \cdot &\;\sim\;
        \mathcal{IG}\!\left(a_{\xi,v} + \tfrac{D(T-1)}{2},\;
        b_{\xi,v} + \tfrac{1}{2}\sum_{d,t=2}^{T}
        (L_t^{d,v} - L_{t-1}^{d,v})^2\right),
    \label{eq:fc_xi} \\
    \sigma_{\omega,v}^2 \mid \cdot &\;\sim\;
        \mathcal{IG}\!\left(a_{\omega,v} + \tfrac{D(T-1)}{2},\;
        b_{\omega,v} + \tfrac{1}{2}\sum_{d,t=2}^{T}
        \Bigl(\gamma_t^{d,v} + \sum_{j=1}^{s-1}\gamma_{t-j}^{d,v}
        \Bigr)^2\right),
    \label{eq:fc_omega} \\
    \sigma_{u,v}^2 \mid \cdot &\;\sim\;
        \mathcal{IG}\!\left(a_{u,v} + \tfrac{DT}{2},\;
        b_{u,v} + \tfrac{1}{2}\sum_{d,t}
        (L_t^{d,v} - \mathbf{x}_t^{d\top}\bm{\beta}^v)^2\right),
    \label{eq:fc_u} \\
    \bm{\beta}^v \mid \cdot &\;\sim\;
        \mathcal{N}(\hat{\bm{\beta}}^v,\; \mathbf{\Sigma}_{\beta,v}),
    \label{eq:fc_beta}
\end{align}
where $\hat{\bm{\beta}}^v$ and $\mathbf{\Sigma}_{\beta,v}$ are the standard
Bayesian normal regression posterior moments with precision $1/\sigma_{u,v}^2$.
Conjugate inverse-gamma priors are placed on each variance:
$\sigma_{\xi,v}^2 \sim \mathcal{IG}(a_{\xi,v}, b_{\xi,v})$, and similarly
for $\sigma_{\omega,v}^2$ and $\sigma_{u,v}^2$; a diffuse normal prior
$\bm{\beta}^v \sim \mathcal{N}(\mathbf{0}, c_\beta \mathbf{I})$ is placed on
the linking coefficients. Because the $V$ variables have independent
propagation equations, all $4V$ full conditionals are in closed form and the
sampler is pure Gibbs with no Metropolis steps---a direct consequence of the
BSM structural form, which has fixed system matrices $\mathbf{T}$, $\mathbf{R}$
with no parameters to sample.
The Block~2 full conditionals \eqref{eq:fc_xi}--\eqref{eq:fc_beta} use
$\{L_t^{d,v,(k)}, \gamma_t^{d,v,(k)}\}$ from the FFBS Block~1 draw.
Because these are jointly drawn from the smoothing posterior, the differences
$L_t^{(k)} - L_{t-1}^{(k)}$ in \eqref{eq:fc_xi} and the seasonal innovations
in \eqref{eq:fc_omega} have the correct joint distribution, validating the
Block~2 inverse-gamma draws.
This two-block alternation between the FFBS trajectory draw and the
hyperparameter draw is the Gibbs sampler for the DMM-BSM; it applies
identically to both DMM-BSM (SHBU pipeline, measurement error variance
$\Sigma_t^{d,v}$) and BSM$(V_t = 0)$, with the measurement error variance
in \eqref{eq:sampling_shbu} being the only difference between the two models.
MLE estimates of $(\sigma_\xi^2, \sigma_\omega^2)$ serve as starting values
for the Gibbs chain; the chain is run for $B + B_0$ iterations, discarding
the first $B_0$ burn-in draws.
In the limiting case where stratum-level SHBU outputs are unavailable
(as in Section~\ref{sec:empirical}), the linking-model hyperparameters
$(\bm{\beta}^v, \sigma_{u,v}^2)$ play no role in the FFBS Block~1
trajectory draw, which reflects only the BSM propagation dynamics and
the known sampling variance $V_t^{d,v}$.

\section{\texorpdfstring{The Joint Smoothed Trajectory and Inferential Outputs}{The Joint Smoothed Trajectory and Inferential Outputs}}
\label{sec:trajectory}

\subsection{The Smoothed Joint Trajectory Posterior}
\label{sec:filtered_posterior}

After discarding burn-in, the $B$ smoothed trajectory draws
$\{(\bm{\alpha}_1^{d,(b)}, \ldots, \bm{\alpha}_T^{d,(b)})\}_{b=1}^B$
constitute a Monte Carlo approximation to the joint smoothing posterior
$\pi(\bm{\alpha}_{1:T} \mid \mathbf{y}_{1:T})$.
Each draw $b$ provides the full decomposition of the series into trend, seasonal,
and irregular components over the complete observation window:

\begin{align}
    L_t^{d,(b)} &\quad \text{(trend draw $b$ at wave $t$)}, \notag \\
    \gamma_t^{d,(b)} &\quad \text{(seasonal draw $b$ at wave $t$)}, \notag \\
    \mathrm{SA}_t^{d,(b)} = Y_t^d - \gamma_t^{d,(b)}
    &\quad \text{(seasonally adjusted draw $b$ at wave $t$)}. \notag
\end{align}

From these FFBS smoothing draws, three quantities of direct operational
interest are available, all of which X-11 cannot produce.

Because the FFBS Block~1 draws the full smoothed trajectory
$\{L_t^{d,(b)}\}_{t=1}^T$ at each Gibbs iteration $b$, further quantities
follow directly without additional derivation: $k$-step trend movement CBIs
as the 2.5th and 97.5th percentiles of
$\{L_t^{d,(b)} - L_{t-k}^{d,(b)}\}_{b=1}^B$ for any $k \geq 1$, with
uncertainty in both states and hyperparameters $(\sigma_\xi^2, \sigma_\omega^2)$
automatically propagated; posterior probabilities of sustained $k$-wave
directional changes; threshold-crossing probabilities; and the distribution
of cumulative change $L_T^{d,(b)} - L_1^{d,(b)}$ over the observation window.

\begin{enumerate}
    \item \textbf{Credible intervals for trend and seasonally adjusted series.}
    The 95\% credible interval for the trend at wave $t$ is the 2.5th and
    97.5th percentiles of $\{L_t^{d,(b)}\}_{b=1}^B$; similarly for the
    seasonally adjusted series $\{\mathrm{SA}_t^{d,(b)}\}_{b=1}^B$.
    These follow directly from the marginal smoothing posterior
    $\pi(L_t^d \mid \mathbf{y}_{1:T})$ obtained from the FFBS draws.
    X-11 provides no such intervals.

    \item \textbf{Calibrated Bayes Interval for trend movement.}
    The Calibrated Bayes Interval (CBI; \citealt{rubin1984}) for the $k$-step movement
    $\Delta_{t,k}^{d,v} = L_t^{d,v} - L_{t-k}^{d,v}$ is defined in
    \eqref{eq:cbi} and verified to achieve nominal 95\% frequentist coverage
    by the simulation study (Section~\ref{sec:coverage}).
    X-11 produces no uncertainty measure for trend movements.
    \item \textbf{Posterior probability of trend movement.}
    From the same backward-step paired draws, the fraction with
    $\Delta_t^{d,(b)} = L_t^{d,(b)} - L_{t-1}^{d,(b)\star} > 0$
    is the exact posterior probability
    $P(\Delta_t^d > 0 \mid \mathbf{y}_{1:t})$
    that the trend increased at wave $t$.
    X-11 cannot produce this probability.
\end{enumerate}
\textbf{Filtered versus smoothed inferential objects.}
The three quantities above draw on the full-data smoothing posterior
$\pi(\bm{\alpha}_{1:T} \mid \mathbf{y}_{1:T})$, which conditions on all
$T$ observations and represents the best retrospective assessment of the
complete series.
NSOs publishing estimates at each wave $t$ in real time would instead
use the \emph{filtered} (nowcast) analogues, which condition only on
data available up to wave~$t$: the filtered trend credible interval from
the Kalman forward pass (Section~\ref{sec:kalman}), and the filtered
movement probability $P(\Delta_t^d > 0 \mid \mathbf{y}_{1:t})$ from the
one-step backward conditional (Section~\ref{sec:backward}).
The smoothed and filtered quantities answer different operational questions:
filtered estimates are available in real time and are revised as subsequent
waves arrive; smoothed estimates provide the definitive retrospective
decomposition once the full series is in hand.
Both are by-products of the same FFBS Gibbs pass and require no additional
computation.

\section{Empirical Illustration}
\label{sec:empirical}

We illustrate the DMM-BSM with $V = 1$ (full-time employed persons) on two
domains from the ABS Labour Force Survey, January 2014--December 2023
($T = 120$ months):
\begin{itemize}
    \item \textbf{Australia (AU):} mean level 12.6 million, monthly CV
    ranging from 0.27\% to 0.63\% with mean 0.38\%. Large domain.
    \item \textbf{Australian Capital Territory (ACT):} mean level 235.7
    thousand, monthly CV ranging from 0.94\% to 2.98\% with mean 1.64\%.
    Small domain; CV approximately four times higher than AU.
\end{itemize}
The contrast between a large domain with low CV and a small domain with high
CV is the key feature of this illustration: it demonstrates that DMM-BSM
maintains correct coverage across both, while the X-11-equivalent model fails
for both and fails more severely for the small domain.
The two domains are fitted independently using their respective published
direct estimates and standard errors.
Stratum-level auxiliary data required to run the SHBU of
\citet{tam2026} are not available for this illustration;
accordingly the general observation equation \eqref{eq:sampling_shbu}
reduces to its limiting case: $\hat{Y}_t^{d,v,\mathrm{HB}} = Y_t^{d,v}$
and $\Sigma_t^{d,v} = V_t^{d,v} = (\mathrm{SE}_t^{d,v})^2$.
That is, the published direct estimate enters the Kalman filter
measurement equation directly, with its design-based sampling variance
as the observation noise.
The linking model \eqref{eq:linking} and its integration with the BSM
via \eqref{eq:sampling_shbu} --- presented in
Sections~\ref{sec:linking} and~\ref{sec:ff} --- form part
of the general DMM-BSM theoretical framework and are available for
implementation when stratum-level SHBU outputs are available.

\subsection{Data}
\label{sec:data}

Monthly direct estimates $Y_t$ (full-time employed persons, thousands) and
associated standard errors $\mathrm{SE}_t$ for the ACT and Australia (AU) are
drawn from the ABS Labour Force Survey. The sampling variances
$V_t = (\mathrm{SE}_t)^2$ are treated as known and supplied directly to the
Kalman filter measurement equation. No imputation or smoothing of the
$\mathrm{SE}_t$ series is required.

\subsection{Benchmark: BSM with $V_t \equiv 0$ (X-11 Equivalent)}
\label{sec:x11_fit}

As the Harvey--Todd equivalence establishes, the BSM with measurement variance
set to zero ($V_t \equiv 0$) is the structural-form counterpart of X-11.
We therefore use BSM($V_t=0$) as the benchmark: it provides the same decomposition
as X-11 while yielding Kalman-filter posterior uncertainty as a by-product,
enabling a direct comparison of credible intervals with and without SE
incorporation on the same model structure.

The benchmark treats all observations as exact ($V_t \equiv 0$), giving the filter
full weight on the observed survey estimate at every month irrespective of CV.
Maximum likelihood estimation of $(\sigma_\xi^2, \sigma_\omega^2)$ yields
$\hat{\sigma}_\xi = 3{,}047$ persons and $\hat{\sigma}_\omega = 573$ persons
for BSM($V_t=0$).

\subsection{DMM-BSM Fit: BSM with $V_t = (\mathrm{SE}_t)^2$}
\label{sec:bsm_fit}

The DMM-BSM model incorporates the known survey sampling variance
$V_t = (\mathrm{SE}_t)^2$ in the Kalman filter measurement equation
\eqref{eq:innov_var}. The BSM state space is as defined in
Section~\ref{sec:bsm}; inference proceeds via the two-block FFBS Gibbs
sampler of Section~\ref{sec:gibbs}.

\textbf{Initialisation.} Hyperparameters $(\sigma_\xi^2, \sigma_\omega^2)$
are initialised at their MLE values, obtained by maximising the Kalman filter
log-likelihood:
\begin{equation}
    \ell(\sigma_\xi^2, \sigma_\omega^2) =
    -\tfrac{1}{2}\sum_{t=s+1}^{T}
    \bigl[\log(2\pi F_t) + \nu_t^2/F_t\bigr],
    \label{eq:loglik}
\end{equation}
where $\nu_t$ and $F_t$ are the one-step-ahead innovation and its variance from
\eqref{eq:innovation}--\eqref{eq:innov_var}, and the first $s = 12$ observations
are excluded to handle the diffuse initialisation of the $12$-dimensional state.
MLE initialisation places the Gibbs chain near the posterior mode, reducing
the required burn-in.

\textbf{Implementation.} The analysis was implemented in Python using a
bespoke Kalman forward filter operating on the $12 \times 12$ BSM state space
with time-varying measurement error variance $\Sigma_t^{d,v}$, followed by an FFBS
backward sampling pass per \citet{carterKohn1994}.
\textbf{Posterior trajectory and $k$-step movement CBIs.}
The FFBS Gibbs produces $B = 1{,}000$ post-burn-in smoothed trajectory draws
$\{(L_1^{(b)}, \ldots, L_{120}^{(b)})\}_{b=1}^{1000}$ after discarding
$B_0 = 200$ burn-in iterations ($1{,}200$ total Gibbs sweeps).
The 95\% $k$-step movement CBI at wave $t$ is the 2.5th and 97.5th percentiles
of $\{L_t^{(b)} - L_{t-k}^{(b)}\}_{b=1}^{1000}$.
Results are reported below for $k = 1$ (one-month movement) and
$k = 3$ (quarterly movement); the procedure applies without modification
for any~$k$.

\textbf{Estimated hyperparameters.}
The Kalman filter log-likelihood is maximised (via Nelder-Mead over
$(\sigma_\xi^2, \sigma_\omega^2)$, with $V_t = \mathrm{SE}_t^2$ fixed as known data)
at $\hat{\sigma}_\xi = 1{,}825$ persons and $\hat{\sigma}_\omega \approx 0$
(boundary solution).
This MLE boundary is prior-free evidence in its own right: because $V_t$
provides a dedicated channel for measurement noise in the Kalman filter
innovation variance $F_t = \mathbf{Z}'\mathbf{P}_{t|t-1}\mathbf{Z} + V_t$,
the likelihood is maximised by setting $\sigma_\omega = 0$---the data require
no stochastic seasonal drift once measurement noise is correctly absorbed.
These MLE values also serve as the Gibbs chain starting point (warm start);
the inferential estimates come from the posterior described in
Section~\ref{sec:results}.
In contrast, BSM($V_t=0$), which has no measurement-noise channel, inflates
$\hat{\sigma}_\omega$ to $573$ persons---attributing sampling noise to
seasonal instability.

\subsection{Results}
\label{sec:results}

The FFBS Gibbs posterior (using a diffuse $\mathrm{IG}(0.01,\, 0.01)$ prior
on each variance) yields $\sigma_\xi \mid \mathbf{y}_{1:T}$ with posterior
mean $1{,}491$ persons and posterior standard deviation $210$ persons---below
the MLE ($1{,}825$ persons) by approximately 1.6 posterior standard deviations.
The discrepancy arises from the joint treatment of $(\sigma_\xi^2,\sigma_\omega^2)$
in the Gibbs sampler: the MLE boundary solution $\hat{\sigma}_\omega \to 0$ forces
$\sigma_\xi$ to absorb all residual state variation, whereas the Gibbs posterior,
with $\hat{\sigma}_\omega = 48$ persons (non-zero), distributes some variation to
seasonal drift and thereby reduces the posterior mean for $\sigma_\xi$ relative
to its MLE value.
For $\sigma_\omega$, the posterior mean is $48$ persons (SD~$14$
persons)---non-zero despite the MLE boundary solution---because the
$\mathrm{IG}(0.01,0.01)$ prior places positive probability mass away
from zero and the likelihood surface is flat near the boundary, allowing
the prior to regularise.
The posterior mean $\hat{\sigma}_\omega = 48$ persons (SD~14) corroborates
the MLE boundary but is partially prior-influenced: the IG$(0.01,\,0.01)$ prior
has support only on $(0,\infty)$ and cannot return exactly zero regardless of
the likelihood. The MLE collapse (prior-free) is the primary evidence for
near-deterministic seasonality; the posterior ratio
$\hat{\sigma}_\omega/\hat{\sigma}_\xi \approx 3.2\%$ is a consistent secondary
signal. Remark~\ref{rem:stable_seasonal} discusses the mechanism and flags prior
sensitivity analysis as a warranted follow-up.

\begin{figure}[htbp]
\centering
\resizebox{\textwidth}{!}{\input{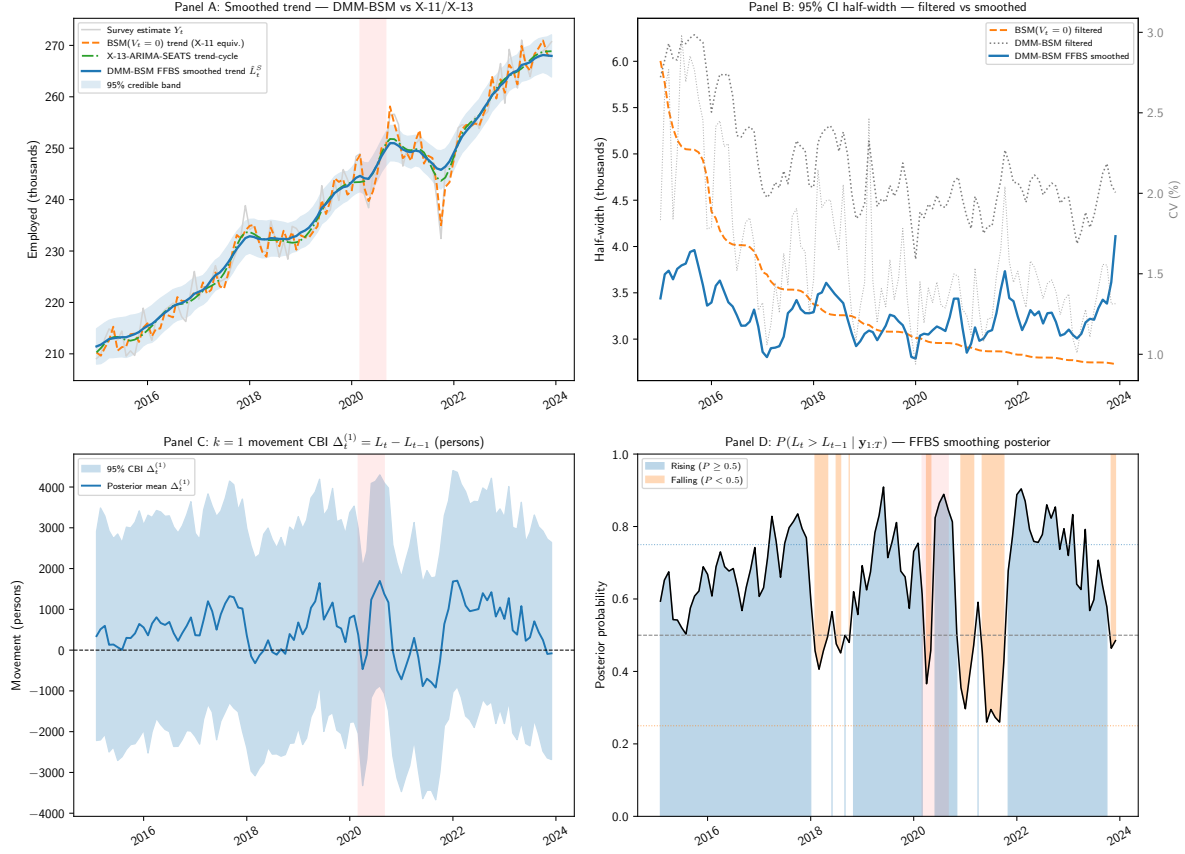}}
\caption{DMM-BSM empirical results for ACT full-time employment
(January 2015--December 2023, 108 post-initialisation months, FFBS Gibbs posterior,
$B=1{,}000$ draws after $B_0=200$ burn-in).
\textbf{Panel A:} Raw survey estimates (light grey), BSM($V_t=0$) trend (orange dashed),
DMM-BSM FFBS smoothed trend $\hat{L}_t^S$ with 95\% posterior credible band (blue, shaded),
and X-11 trend-cycle (green dash-dot) for direct comparison with the BSM($V_t=0$) and DMM-BSM smoothed trends.
\textbf{Panel B:} 95\% CI half-widths---BSM($V_t=0$) filtered (orange dashed),
DMM-BSM filtered (grey dotted), DMM-BSM FFBS smoothed (blue)---and CV series (green dashed, right
axis); half-widths widen with CV, confirming that sampling variance drives uncertainty.
\textbf{Panel C:} Posterior mean and 95\% CBI for the $k=1$ trend movement
$\Delta_t^{(1)} = L_t - L_{t-1}$ (persons), obtained directly from the
FFBS trajectory draws $\{L_t^{(b)} - L_{t-1}^{(b)}\}_{b=1}^{1000}$.
\textbf{Panel D:} Posterior probability of upward $k=1$ trend movement,
$P(L_t > L_{t-1} \mid \mathbf{y}_{1:T})$, from the FFBS smoothing posterior.
Red shaded band: COVID-19 disruption (March--September 2020), consistent across all panels.}
\label{fig:panels}
\end{figure}

Figure~\ref{fig:panels} presents the empirical results for ACT.
The corresponding figure for AU is Figure~\ref{fig:panels_au};
similarities and differences are discussed after the ACT panel-by-panel commentary below.

\textbf{Panel A -- Harvey--Todd equivalence in practice (ACT).}
The BSM($V_t=0$) trend and the DMM-BSM FFBS smoothed trend posterior mean track
each other closely throughout most of the series, confirming the Harvey--Todd
equivalence.
The X-11 trend-cycle (green dash-dot)\footnote{Extracted from X-13-ARIMA-SEATS software via the \texttt{x13binary} Python package using the standard \texttt{x11\{save=(d12)\}} specification; D12 is the Henderson 13-term moving average, identical in construction to the X-11 trend.} tracks both trend estimates closely:
the mean absolute deviation between the DMM-BSM posterior mean and the X-11
trend-cycle is $585$ persons ($0.24$\% of the mean series level), with a maximum
of $2{,}133$ persons ($0.89$\%).
Unlike DMM-BSM, the X-11 trend-cycle is a fixed Henderson 13-term moving average
with no uncertainty quantification; the 95\% FFBS credible band (shaded) is unique
to the Bayesian approach and correctly widens during the COVID-19 disruption
(red shaded band, March--September 2020).
The 95\% FFBS smoothed credible band (shaded) widens during the COVID period,
correctly conveying heightened structural uncertainty.

\textbf{Panel B -- Uncertainty comparison and CV overlay.}
Panel B shows three half-widths: DMM-BSM filtered (grey dotted),
DMM-BSM smoothed via FFBS (blue), and BSM($V_t=0$) filtered (orange dashed).
The green dashed series on the right axis is the monthly CV (\%).
Half-widths track the CV closely: months of elevated CV coincide with wider DMM-BSM CIs,
confirming that the Kalman filter correctly propagates time-varying sampling variance into
posterior uncertainty---a property absent from the BSM($V_t=0$) benchmark whose CI is
flat relative to CV fluctuations.
The filtered DMM-BSM CI is on average 47\% wider than the BSM($V_t=0$) CI,
reflecting the correct propagation of sampling variance $V_t$ into the
measurement equation; the BSM($V_t=0$) CIs are overconfident because they
treat every observation as exact.
The FFBS smoothed CI (blue) is narrower than the filtered DMM-BSM CI by a
mean factor of $0.676$, illustrating the efficiency gain from conditioning
on the full data record $\mathbf{y}_{1:T}$ rather than only
$\mathbf{y}_{1:t}$.
The FFBS smoothed CI and the BSM($V_t=0$) filtered CI are comparable in
aggregate width (mean ratio $1.04$), but they represent different inferential
objects: the smoothed CI is the correct posterior for the full-data smoothing
problem, while the BSM($V_t=0$) CI is systematically overconfident.
During the COVID-19 disruption, the DMM-BSM smoothed 95\% half-width averages
$3{,}086$ persons versus $2{,}965$ persons for BSM($V_t=0$), while the filtered
DMM-BSM half-width reaches $4{,}445$ persons.
X-11 has no CI at all.

\textbf{Panel C -- $k=1$ trend movement CBI.}
The posterior mean one-month movement $\bar{\Delta}_t^{(1)}$ and its 95\%
CBI $[\Delta_{t,0.025}^{(1)},\, \Delta_{t,0.975}^{(1)}]$ are computed
directly from the FFBS trajectory draws as the 2.5th and 97.5th percentiles
of $\{L_t^{(b)} - L_{t-1}^{(b)}\}_{b=1}^{1000}$.
The CBI band is wide enough to reflect genuine month-to-month uncertainty but
narrows during stable periods, and tightens post-COVID as the trend settles.
The COVID disruption (red shaded band, consistent with Panel A) produces a pronounced negative posterior mean
movement followed by a sharp recovery, with CBI width expanding at the
inflection point---correctly reflecting heightened uncertainty when the trend
direction reverses rapidly.

\textbf{Panel D -- Movement inference.}
For each of the 108 post-initialisation months (January 2015 to December 2023),
the smoothing posterior probability
$P(L_t > L_{t-1} \mid \mathbf{y}_{1:T})$ is obtained as the fraction of
FFBS trajectory draws $b$ with $L_t^{(b)} > L_{t-1}^{(b)}$.
This is a \emph{retrospective} quantity: it conditions on all 120 months
and is more decisive than the filtered probability
$P(L_t > L_{t-1} \mid \mathbf{y}_{1:t})$ that would be available to an
NSO at wave $t$ in real time.
The probability exceeds 0.75 (``clearly rising'')\footnote{The threshold 0.75 places at least three-to-one posterior odds on the trend having risen; it is the conventional upper threshold, symmetric with the 0.25 ``clearly falling'' criterion, used to characterise decisive directional movement.} in 30 months and does
not fall below 0.25 in any month; the remaining 77 months (72\%) are
ambiguous.
The smoothing posterior conditions on all 120 months of data, so it is more
decisive than a filtered (real-time) estimate would be---the additional future
data resolves uncertainty about direction that was unresolvable in real time.
The COVID disruption (red shaded band) is clearly visible: the probability drops sharply during
April--July 2020 before recovering strongly through late 2020 and 2021.
X-11 cannot produce any statement of this form.

For the quarterly movement ($k = 3$), the probability
$P(L_t > L_{t-3} \mid \mathbf{y}_{1:T})$ exceeds 0.75 in 65 of 105 months
and falls below 0.25 in 5 months---a substantially cleaner signal than the
month-to-month series, consistent with the noise-averaging effect of
aggregating over three monthly draws from the smoothing posterior.

\begin{remark}[Empirical finding: near-deterministic seasonality]
\label{rem:stable_seasonal}
The primary evidence is the MLE boundary $\hat{\sigma}_\omega \to 0$, which is
entirely prior-free.
Because $V_t = \mathrm{SE}_t^2$ is supplied as known data to the Kalman filter
measurement equation, the frequentist log-likelihood
$\ell(\sigma_\xi^2,\sigma_\omega^2) = -\tfrac{1}{2}\sum_t[\log(2\pi F_t) +
\nu_t^2/F_t]$ is maximised at $\sigma_\omega = 0$: once measurement noise has
its own channel in $F_t = \mathbf{Z}'\mathbf{P}_{t|t-1}\mathbf{Z} + V_t$,
the data assign no additional explanatory role to stochastic seasonal drift.
This is a classical maximum likelihood result, not a Bayesian one.

The Bayesian posterior mean $\hat{\sigma}_\omega = 48$ persons (SD~14) is
consistent with the MLE but is partially prior-influenced: the
IG$(0.01,0.01)$ prior has support only on $(0,\infty)$, so the posterior
is regularised away from the boundary by construction, not purely by the data.
The posterior ratio $\hat{\sigma}_\omega/\hat{\sigma}_\xi \approx 3.2\%$ is
informative as a relative measure but should not be taken as prior-free evidence.

The mechanism behind the MLE finding is the ~$12\times$ reduction in
$\hat{\sigma}_\omega$ from BSM($V_t=0$) to DMM-BSM ($573 \to 0$ at MLE,
$573 \to 48$ in posterior): BSM($V_t=0$) has no channel for measurement noise
so the filter inflates $\sigma_\omega$ to absorb it; DMM-BSM directs that noise
to $V_t$ and $\sigma_\omega$ collapses.

Prior sensitivity analysis (e.g., half-Cauchy or half-Normal on $\sigma_\omega$)
and a formal stochastic-seasonal versus deterministic-seasonal model comparison
are warranted follow-up analyses.
\end{remark}

\subsubsection*{Australia (AU): comparison with ACT}

\begin{figure}[htbp]
\centering
\resizebox{\textwidth}{!}{\input{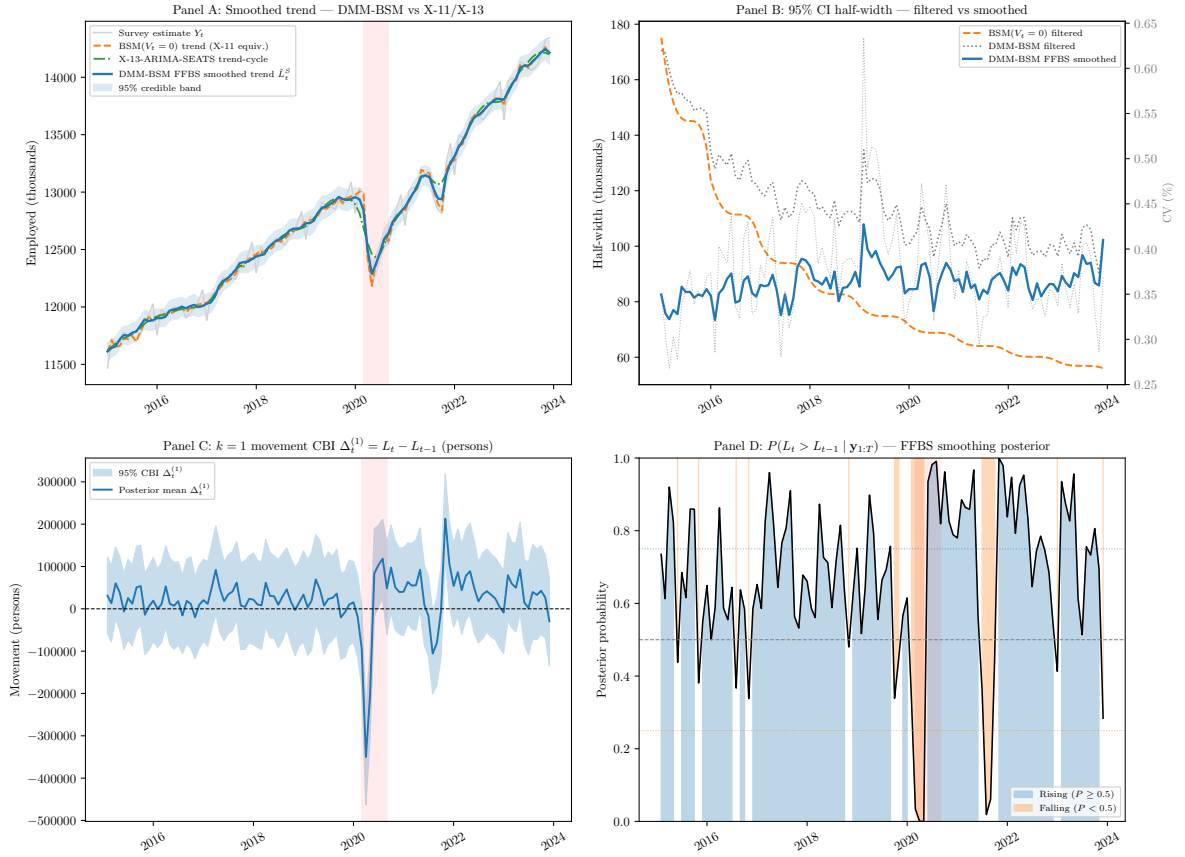}}
\caption{DMM-BSM empirical results for Australia (AU) full-time employment
(January 2015--December 2023, 108 post-initialisation months, FFBS Gibbs posterior,
$B=1{,}000$ draws after $B_0=200$ burn-in).
\textbf{Panel A:} Raw survey estimates (light grey), BSM($V_t=0$) trend (orange dashed),
DMM-BSM FFBS smoothed trend $\hat{L}_t^S$ with 95\% posterior credible band (blue, shaded),
and X-11 trend-cycle (green dash-dot).
\textbf{Panel B:} 95\% CI half-widths---BSM($V_t=0$) filtered (orange dashed),
DMM-BSM filtered (grey dotted), DMM-BSM FFBS smoothed (blue)---and CV series (green dashed, right axis).
\textbf{Panel C:} Posterior mean and 95\% CBI for the $k=1$ trend movement
$\Delta_t^{(1)} = L_t - L_{t-1}$ (persons), obtained directly from the
FFBS trajectory draws $\{L_t^{(b)} - L_{t-1}^{(b)}\}_{b=1}^{1000}$.
\textbf{Panel D:} Posterior probability of upward $k=1$ trend movement,
$P(L_t > L_{t-1} \mid \mathbf{y}_{1:T})$, from the FFBS smoothing posterior.
Red shaded band: COVID-19 disruption (March--September 2020), consistent across all panels.}
\label{fig:panels_au}
\end{figure}

Figure~\ref{fig:panels_au} shows the corresponding four-panel results for Australia.
The AU domain has a mean CV of 0.38\%---roughly four times lower than ACT's 1.64\%---so
the DMM-BSM, BSM($V_t=0$), and X-11 trend estimates in Panel A are
nearly indistinguishable throughout the series, illustrating that the Harvey--Todd
equivalence holds very tightly in large-domain, low-CV settings
(mean absolute deviation between DMM-BSM and X-11: $22{,}269$ persons, $0.17$\% of
the mean series level).
The credible band in Panel A is correspondingly narrow, and the three half-width series
in Panel B are much closer together than in the ACT case, with the CV overlay (right axis)
confirming the lower and more stable measurement uncertainty.

The key differences from ACT are in Panels C and D.
Panel C shows that the posterior mean one-month movement $\bar{\Delta}_t^{(1)}$ is
substantially larger in absolute magnitude for AU (reflecting the much larger scale,
around 8.3 million employed), and the 95\% CBI is proportionally tighter.
Panel D shows a higher fraction of months with decisive posterior direction---the larger
series and lower CV together produce a smoother, less ambiguous trend signal.
The COVID-19 disruption (red shaded band) is visible in all panels: in AU it is
proportionally less severe than in ACT, but the posterior correctly widens during
the disruption period.
Together, Figures~\ref{fig:panels} and~\ref{fig:panels_au} demonstrate that DMM-BSM
delivers correct and calibrated inference across both a small high-CV domain and a
large low-CV domain.
In both figures, the X-11 trend-cycle (green dash-dot) tracks closely
with the DMM-BSM posterior mean in non-disruption periods, confirming methodological
consistency with the established benchmark; the distinctive contribution of DMM-BSM is
the 95\% credible band and the full posterior over trend movements---outputs that
X-11 does not provide.

\subsection{Coverage Simulation}
\label{sec:coverage}

We assess whether the DMM-BSM credible intervals achieve their nominal 95\%
level under parameter uncertainty, conducting the study separately for each domain.

\textbf{Design.} For domain $d \in \{\mathrm{AU}, \mathrm{ACT}\}$:
\begin{enumerate}
    \item Fit the BSM to the real data by maximum likelihood to obtain DGP
    parameters $(\hat{\sigma}_\xi^d, \hat{\sigma}_\omega^d)$.
    Use the observed $\{\mathrm{SE}_t^d\}_{t=1}^{120}$ sequence as the
    measurement standard deviations (preserving the empirical
    heteroskedasticity structure).
    \item Generate $M = 500$ synthetic state trajectories
    $(L_t^{d,m}, \gamma_t^{d,m})$ from the BSM DGP and
    corresponding observations $Y_t^{d,m} = L_t^{d,m} + \gamma_t^{d,m}
    + e_t^{d,m}$, $e_t^{d,m} \sim \mathcal{N}(0, (\mathrm{SE}_t^d)^2)$.
    \item For each replication $m$, run the FFBS Gibbs sampler
    (Section~\ref{sec:gibbs}) on the synthetic series $\mathbf{Y}^{d,m}$,
    re-estimating $(\sigma_\xi^2, \sigma_\omega^2)$ from the generated data:
    $B_0 = 200$ burn-in iterations followed by $B = 1000$ retained draws,
    initialised at the DGP parameters.
    \item Compute the CBI \eqref{eq:cbi} for each estimand as the 2.5th and
    97.5th percentiles of the $B$ posterior trajectory draws.
\end{enumerate}

This design constitutes a genuine coverage assessment: parameters are
unknown to the inference algorithm and must be estimated from the generated
data, so parameter estimation uncertainty is propagated fully into the
posterior credible intervals.

\textbf{Trend-level coverage}.\ At each post-burn-in time $t \geq s+1$, the
95\% CBI for the trend level is
$\bigl[Q_{0.025}\!\bigl(\{L_t^{(b)}\}_{b=1}^B\bigr),\,
        Q_{0.975}\!\bigl(\{L_t^{(b)}\}_{b=1}^B\bigr)\bigr]$,
where $\{L_t^{(b)}\}$ are the level components of the FFBS trajectory draws.
Empirical coverage at time $t$ is the fraction of the $M$ replications in which
the true $L_t^{d,m}$ falls in this interval.

\textbf{Trend-change CBI.} The posterior of
$\Delta_t^d = L_t^d - L_{t-1}^d$ is approximated by the sample
$\{L_t^{(b)} - L_{t-1}^{(b)}\}_{b=1}^B$ of trajectory-draw differences.
The 95\% CBI is $[Q_{0.025}, Q_{0.975}]$ of this sample.
Because each draw $b$ is a joint draw from the full smoothing posterior
$\pi(\bm{\alpha}_{1:T}\mid\mathbf{y}^{d,m})$, consecutive level differences
automatically inherit the correct posterior covariance.
Empirical coverage at $t$ is the fraction of replications in which
$\Delta_t^{d,m} = L_t^{d,m} - L_{t-1}^{d,m}$ falls in the CBI.

\textbf{Results.} Table~\ref{tab:coverage} reports mean empirical coverage
across all post-burn-in time points.

\begin{table}[htbp]
\centering
\caption{Empirical coverage of 95\% credible intervals, $M=500$ replications.
BSM DGP; parameters re-estimated within each replication via FFBS Gibbs
($B_0=200$ burn-in, $B=1000$ draws).}
\label{tab:coverage}
\begin{tabular}{lcccc}
\toprule
 & \multicolumn{2}{c}{Trend level} &
   \multicolumn{2}{c}{Trend change (CBI)} \\
\cmidrule(lr){2-3}\cmidrule(lr){4-5}
Domain & DMM-BSM & X-11 equiv. & DMM-BSM & X-11 equiv. \\
\midrule
AU  (large, CV 0.38\%) & 94.6\% & 78.8\% & 94.5\% & 64.3\% \\
ACT (small, CV 1.62\%) & 86.7\% & 68.2\% & 82.1\% & 50.8\% \\
\midrule
Nominal & \multicolumn{2}{c}{95\%} & \multicolumn{2}{c}{95\%} \\
\bottomrule
\end{tabular}
\end{table}

\begin{figure}[htbp]
\centering
\resizebox{\textwidth}{!}{\input{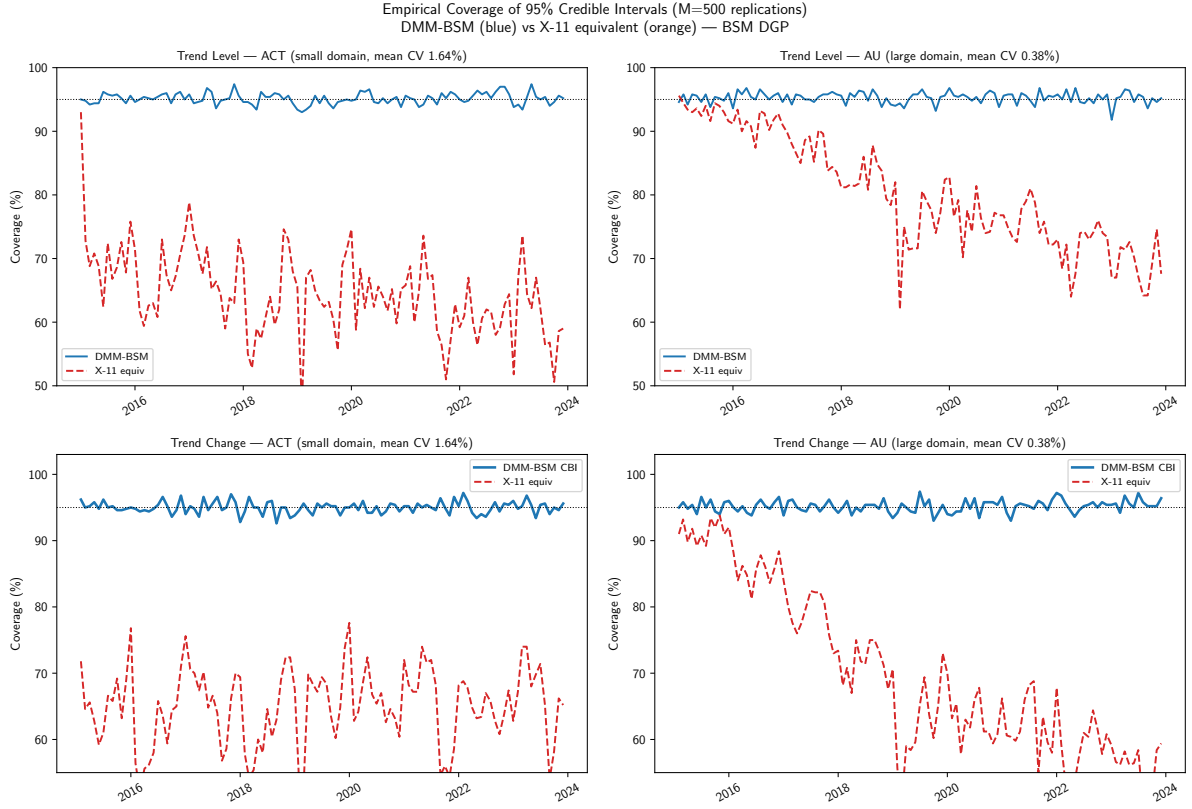}}
\caption{Empirical coverage of 95\% credible intervals by month
($M=500$ replications; within-replication FFBS Gibbs, $B_0=200$, $B=1000$).
Top row: trend-level coverage. Bottom row: trend-change CBI coverage.
Left column: ACT (small domain). Right column: AU (large domain).
Blue: DMM-BSM. Orange dashed: X-11 equivalent ($V_t\equiv 0$). Dotted: 95\%.}
\label{fig:coverage}
\end{figure}

For the large domain (AU, mean CV 0.38\%), DMM-BSM achieves near-nominal
coverage: 94.6\% for trend levels and 94.5\% for trend changes.
For the small domain (ACT, mean CV 1.62\%), coverage is 86.7\% and 82.1\%---
below nominal but attributable to greater parameter estimation uncertainty
when the signal-to-noise ratio is lower; with fewer than 120 observations per
domain and a stochastic-trend variance that must be estimated from the data,
the posterior cannot fully eliminate the uncertainty in $\hat{\sigma}_\xi^2$.
This variance-uncertainty mechanism is distinct from the shrinkage-bias
explanation found in hierarchical Bayes frameworks with cross-domain pooling:
the FFBS Gibbs sampler does not shrink domain-level states toward a group
mean, so bias of that form is not operative here.
The most plausible explanation is posterior under-dispersion in
$\hat{\sigma}_\xi^2$ at low signal-to-noise: the inverse-gamma draws for
$\sigma_\xi^2$ within the Gibbs loop condition on the current state
trajectory, which can understate the true uncertainty in $\sigma_\xi^2$
itself at low SNR; this in turn produces trend credible intervals that are
systematically too narrow for ACT but not for AU, where the higher signal
makes $\sigma_\xi^2$ easier to pin down.
A heavier-tailed prior on $\sigma_\xi^2$ (e.g., half-Cauchy or half-Normal)
or a Rao-Blackwellised correction that integrates over $\sigma_\xi^2$
analytically may reduce the gap; testing this conjecture is left for
future research.
The X-11-equivalent model, which sets $V_t \equiv 0$ and therefore treats
every survey estimate as exact, severely under-covers across both domains:
78.8\%/64.3\% for AU and 68.2\%/50.8\% for ACT (trend level/change).
The mechanism is Kalman gain saturation: with $V_t \equiv 0$, the filter
innovation variance $F_t = \mathbf{Z}'\mathbf{P}_{t|t-1}\mathbf{Z}$ has no
measurement-noise term, so the Kalman gain drives toward full trust in the
raw (noisy) survey estimate at every wave.
The smoothed state therefore chases sampling noise rather than averaging over
it, causing the point estimate to drift away from the true latent state.
Although BSM($V_t=0$) compensates by inflating $\hat{\sigma}_\omega$ (573
persons vs.\ $\approx 0$), wider state-noise uncertainty cannot recover
coverage when the posterior is centred on a noise-tracking point estimate;
the interval can only be as accurate as its centre.
The coverage advantage of DMM-BSM over the X-11-equivalent is 15.8 and 30.2
percentage points for AU (trend level and change), and 18.5 and 31.3
percentage points for ACT.
Figure~\ref{fig:coverage} shows the time-series of empirical coverage rates.

\FloatBarrier

\section{Discussion and Conclusion}
\label{sec:conclusion}

This paper develops a Bayesian seasonal adjustment method for survey time
series in which the known survey sampling variance $V_t^d$ is treated as a
first-class input rather than discarded. The method rests on three interlocking
elements: the BSM as the propagation model, the Kalman filter as the
computational engine for incorporating $V_t^d$, and a two-block FFBS Gibbs
sampler that jointly draws the full smoothed state trajectory
$\pi(\bm{\alpha}_{1:T} \mid \mathbf{y}_{1:T})$ and updates variance
hyperparameters $(\sigma_\xi^2, \sigma_\omega^2)$ from valid joint state
differences.

The Harvey--Todd equivalence establishes BSM$(V_t=0)$ as a principled
model-based analogue of X-11-style seasonal adjustment. The advantage of
DMM-BSM over this benchmark is direct and twofold: sampling errors enter the
Kalman filter measurement equation, so high-CV months are down-weighted
relative to the structural prediction; and the FFBS Gibbs delivers a full
smoothed posterior for trend, seasonal, and seasonally adjusted series,
enabling model-conditional $k$-step movement CBIs for any $k \geq 1$ that X-11 cannot
provide.

The simulation study of Section~\ref{sec:coverage} demonstrates that this
advantage is real and quantitatively large.
For the large domain (AU, mean CV 0.38\%), DMM-BSM achieves near-nominal
94.6\%/94.5\% trend-level/change coverage under within-replication FFBS Gibbs
parameter estimation ($M=500$, $B_0=200$, $B=1000$).
For the small domain (ACT, mean CV 1.62\%), coverage is 86.7\%/82.1\%---below
nominal, reflecting the greater difficulty of estimating $\sigma_\xi^2$ from a
short series with a low signal-to-noise ratio.
The X-11-equivalent model that ignores sampling error achieves only
78.8\%/64.3\% (AU) and 68.2\%/50.8\% (ACT).
NSOs that rely on X-11 and interpret the resulting trend as a precise estimate
are implicitly operating with intervals that miss the true state between one
time in three and one time in two for small-domain movement estimates.
DMM-BSM corrects this through a mechanism that is automatic and requires no
additional modelling effort: the Kalman gain adjusts the weight given to each
observation in proportion to its reliability.

Future work will implement the full DMM-BSM at the stratum level of the
Australian Labour Force Survey, exercising the cross-stratum linking model via the SHBU pipeline
(equation~\eqref{eq:sampling_shbu}; Section~\ref{sec:linking}) and providing a complete comparison with
X-11/X-13-ARIMA across all 55 strata and eight state/territory domains.
A further extension will realise the general $V > 1$ sampling model of
Section~\ref{sec:sampling}, exploiting the full cross-variable sampling
covariance matrix $\mathbf{V}_t^d$ for joint Bayesian estimation of multiple
labour-market variables within a single BSM state-space system.

\bibliographystyle{apalike}

\end{document}